# On costly signalling in DAOs: A research agenda

Darcy W. E. Allen[1], Jason Potts[1], Julian Waters-Lynch[1] and Max Parasol[1]

[1] RMIT Blockchain Innovation Hub, RMIT University
`jason.potts@rmit.edu.au`

**Abstract.** Decentralised Autonomous Organisations (DAOs) are a new type of digital organisation that uses blockchain infrastructure (e.g. smart contracts, tokens) to coordinate a group of people around a shared mission. Like all organisations, DAOs must attract sources of funding and other resources, and discover and retain a talented community and workforce. To do this, they must signal their true quality. Yet the characteristics of the environment that DAOs operate in — pseudonymous actors, global scale, permissionless entry and exit — makes this difficult. We apply costly signalling theory to explore the information asymmetry problem in DAOs and some of the strategies (behaviours and investments) and institutional solutions (including better signalling mechanisms) that have evolved to solve this problem.

**Keywords:** Decentralised Autonomous Organisations, DAO, Cryptoeconomics, Costly Signalling, DAO Governance

## 1    Introduction

Blockchain and smart contract technologies have made possible a new type of organisation – the Decentralised Autonomous Organisation, or DAO (Buterin 2013, 2014; Voshmgir 2017; Hassan and De Filippi 2021). The benefits that DAOs have over traditional centralised, hierarchical organisations are in lowering transaction costs and reducing agency problems due to greater transparency, predictability and assurance. DAOs often offload organisational functions to software code and use blockchain consensus and smart contracts to enable these operations to be trusted in an open, permissionless context. These features make DAOs competitive new organisational forms for a digital economy. Both the open-source nature of DAOs and the emergence of toolkits (Allen and Potts 2023) means that these frontier organisations are also increasingly low cost to spin up and to experiment with. But this novelty also imposes costs of not being adapted to existing institutional orders and the solutions they provide.

In this paper we focus on the DAO signalling problem, including how DAOs deal with the asymmetric information about quality in different contexts (e.g. seeking investment, sourcing talented workers). We examine the emergent strategies and solutions that DAOs have evolved, such as new tooling, to help provide these signals. Our core conjecture is that the process of developing new signalling mechanisms is more expansive than currently appreciated and that there are multiple directions over which signals are being sent (and invested in). These include workers signalling to DAOs their hidden qualities in a competitive labour market, and DAOs signalling their true type to counterparties (e.g. members).

These problems emerge because of asymmetric private information about quality. Overcoming these problems is necessary for engaging the community, hiring staff and contractors, attracting investors and managing public perception of legitimacy, particularly for regulatory and governance oversight. DAOs that ameliorate these problems will have a competitive advantage. Such problems are consequences of noisy environments that create strategic opportunities for some players to misrepresent their true type. Players and organisations can engage in 'cheap talk', claiming to be high quality when their true state is low quality (Farrell and Rabin 1996). These asymmetric information problems occur when some agents have private information that others do not. It becomes difficult to evaluate the quality of counterparties, leading to problems of economic choice and coordination (Stiglitz, 1985, 2002; Löfgren et al. 2002). If I have a high-quality product or organisation, but that information is not known to you, or you distrust my claims about quality, that will affect your decisions about offers made to purchase my product or join my organisation. In this case, product markets and factor markets can fail, possibly completely (Akerlof 1970).

There are two classes of solution to the asymmetric information problem: (1) *screening* mechanisms that focus on the principal designing an incentive for others to reveal true information (see Stiglitz, 1975); and (2) *costly signalling* mechanisms where the principal provides a costly signal to reveal to others their true type (e.g. Spence 1973). The former approach, screening, is when the principal in a transaction designs a high-powered incentive for agents to invest in revealing true information. We can see screening around deductibles in insurance where parties are incentivized to reveal their true risk level (Rothschild and Stiglitz 1976), or by revealing personal qualities to gain admission to a coveted organisation. A DAO could screen for desirable qualities (i.e. incentivising revelation of private information) by the various options given to join particular groups (e.g. Telegram groups). One way that a DAO could screen for technical competence, for instance, is by varying how difficult these are to interact with.

Our focus in this paper is on costly signalling mechanisms in the context of DAOs. A costly signal involves the principal sending an honest signal to the market about their true type. For instance, if I am a high-type DAO (i.e. high quality), how do I reliably communicate that information to counterparties, so that they might factor that information into their decision-making about how to interact with me? How can I distinguish my organisations from low-type DAOs? How do I avoid the 'pooling equilibrium' and the reduced pay-offs associated with being confused with lower quality competitors? Costly signalling mechanisms are potential solutions to these problems, and we are beginning to see them emerge in the DAO ecosystem.

Almost everything that engages in strategic interaction in the world has this problem of reliably signalling quality (Zahavi 1975, Grafen 1990, Smith and Bird 2000, Henrich 2009, Pentland 2008). These signals must be communicated in multiple directions, including from organisations to investors and from workers to potential employers. Over hundreds of years, modern industrial capitalism has evolved and adapted many specific institutional solutions to this problem that take advantage of costly signalling solutions. Such mechanisms include licensing regimes, regulation, financial markets, venture capital, hiring consultants, and underwriters (Leland and Pyle 1977, Ross 1977, Bhattacharya, 1979, 1980, Carter et al. 1998, Kirmani and Rao 2000, Davila et al. 2003, Bergh and Gibbons 2011, Janssen and Roy 2015). Yet the technological innovations that make DAOs possible also mitigate or even eliminate these evolved institutional solutions. The mechanisms must work in a permissionless, global and pseudononymous environment. Today DAOs face the perennial and fundamental problem of communicating their true type (a difficult to observe but critical to know quality) to counterparties, including investors, DAO members and regulators.

While previous work applying signalling theory to blockchain has examined the effectiveness of cheap signals in token offerings (Ante and Fiedler 2020), we focus on the potential costly strategic actions and investments that DAOs make in order to signal quality, and explore the type of 'separating equilibria' they induce (Bergstrom et al.

2002). We also study how new institutional forms are evolving to facilitate signalling mechanisms, the supply of entrepreneurial solutions, and the collective action problems involved.

## 2 Signalling Theory

A costly signal is a solution to the asymmetric information problem. Spence (1973) first proposed this mechanism in the context of labour markets – a higher education degree is a costly signal of underlying but directly unobservable quality of the job market candidate. In parallel, biologists were developing a similar theory about expensive evolutionary adaptations in sexual selection, such as peacock tails and deer antlers (Zahavi 1975). The evolutionary mechanism is the same in both cases – how to reliably send an honest reliable signal of an important but unobservable quality (e.g. the genetic quality of a mate, the toxicity of a prey animal, the true capabilities of a prospective employee) to specific targets whose behaviour is conditional upon true information (e.g. mates, predators or employers) and to do so in a way that those without the quality (low quality mates, delicious prey, slack employees) would find it too costly to send the signal. This aspect is called the separating equilibrium (Bergstrom et al. 2002), and it is what enables the signal to be cheaply observed, but expensive to produce if you lack the underlying quality. Costly signals facilitate efficient discrimination and avoid inefficient pooling equilibria.

Costly signalling is a theory of strategic communication in a world of imperfect information that works by performing an action (investing resources to send a signal) that is costly to all who send the signal, but differentially costly depending on whether the sender of the signal does or does not have the underlying quality (Connelly et al. 2011). High-types, who have the underlying quality, find it relatively cheaper to send the signal than low-types, for whom the signal is relatively more expensive. A separating equilibrium occurs when the relative cost is such that low types don't signal at all. For example, a luxurious, long and heavy tail is costly for a peacock with good health and genes, as it makes evading predators more difficult and requires more energy and strength to sustain. But it is prohibitively and perhaps lethally expensive for a weaker peacock of poor genetic quality. The peahen, choosing a prospective mate, can quickly ascertain quality by simply observing the peacock's tail, which, if of high-type, he will proudly display. The same argument applies to other costly signals, such as degrees from top universities (expensive, wasteful and difficult for everyone, but relatively more so for weaker students, Caplan 2018), bright aposematic displays on toxic animals (making them more visible to predators, Sherratt 2002), issuing generous dividends (which only a well-run, profitable firm can afford, Ross 1977), and so on.

Costly signals are investments in sending true information that distinguishes them from false signals ('cheap talk', Farrell and Rabin 1996) because of signal cost. Costly signals work when they are difficult to fake or bluff, and they will evolve to increase in cost until that condition holds. Indeed, the equilibrium level of costliness occurs when those seeking to send a fake signal find it prohibitively costly (i.e. no longer rational) to do so. Note the signal need have no intrinsic benefit; indeed, it works best when actively wasteful or harmful to the carrier because weaker agents cannot stand the cost of sending it.

Sending costly signals requires some kind of signalling mechanism. In biology, that works through genes and phenotypic expressions (of tails, antlers, calls, behaviours, etc). In the social and economic realm, costly signalling occurs through investments by agents (e.g. time and effort to get a higher degree) but the effectiveness of these investments in signals relies on the entire education system, including certification mechanisms, quality control and development of institutional norms about their meaning. These institutional structures supply the signalling mechanisms that co-evolve with demand by high-type agents to send costly signals. The origin of these signalling mechanisms is unexplored in the signalling literature, with their existence either assumed, or treated as an outcome of cooperation (Gintis et al. 2001). But it is reasonable to suppose that these mechanisms are entrepreneurially supplied (examples in section 4 below) and form economic infrastructure to facilitate competition in the DAO ecosystem (see Jacobides et al. 2018).

## 3  DAOs and their information problems

DAOs are a new type of digital organisation based on smart contracts in which groups of people use blockchain infrastructure to coordinate around some shared objective (Vergne 2020, Hasan and de Filippi 2021, Wright 2021). Smart contracts are digital agreements to automate processes and decisions typically handled by humans. Santana and Albareda (2022: 1-2) define DAOs as "...blockchain-based organizations fed by virtual open networks of contributors (investors in cryptocurrencies). Their governance and management are decentralized without central control and are built on automated rules encoded in smart contracts stored and executed in blockchains. This structure enables peers to work autonomously based on a system of on-chain (machine consensus) and off-chain (voting rights) mechanisms of governance that support community decision-making and drive distributed trust among peers."

The number of DAOs in existence today is around 50,000, although only about 2500 have more than one million in assets. Notable examples such as layer2 protocols (e.g. Optimism, Arbitrum), Gnosis (DAO services, e.g. multisig wallet), MakerDAO (a decentralised stablecoin protocol), Gitcoin (open source software funding), and Uniswap (a decentralised exchange protocol). Many early DAOs have failed, some spectacularly so (DuPont 2017). DAOs appeal to people who are dissatisfied with traditional forms of work and organisations, including those accustomed to open-source software governance, as well as those seeking more egalitarian, transparent and collaborative ways of working with a stronger sense of community.

There are various taxonomies of DAO types and their architectures (Ziegler and Welpe 2022). Wright (2021) distinguishes between algorithmic DAOs, which defer entirely to software, and participatory DAOs, which also have human exert control through governance tokens. Wang et al., (2019) conceptualised DAOs as having five layers of system components including the protocol, incentive mechanisms and a goal. In a short time, DAOs have been applied to a large range of use cases, including managing blockchain protocols, pooling funds to buy objects or to organise other club-like behaviour, and acting as a mechanism for foundation governance. Yet all DAOs must attract, motivate, incentivise and coordinate participants to undertake work tasks to further the goals of the DAO, and in turn, those candidates must be evaluated.

Because of their context, DAOs have complex work arrangements (Atherton et al. 2020, Ilyushina and MacDonald 2022). An important difference between DAOs and conventional organisations is their approach to finance and compensation, where they often finance work through their own native cryptocurrency tokens. Work in DAOs can be contract-based, team-based, time-based, project-based, task-based, or contribution-based (Rennie and Potts 2024). Different strategies have different incentives by treating workers more like suppliers of contract labour, or more like equity partners. Wage-based compensation transfers risk and contracting cost (including monitoring) to the organisation, so has lower expected value to the employee, but also low variance, which is often highly valued. Contract bidding for tasks is a screening mechanism, intended to get suppliers to reveal their true cost of supply. Equity-based compensation incentivises early-stage employees to work harder for less pay due to the possibility of a sizable pay-out in the future, usually after a vesting period to align incentives to commit to the team. Paying workers with native crypto tokens has advantages for DAOs, as they can mint tokens without relying on external funding. Yet projects can issue tokens before they have a viable business model, which has led to concerns about scams and vaporware (Howell et al., 2020). Token compensation falls somewhere between cash and equity. Tokens are more liquid than startup equity when convertible on a crypto exchange, but value is often volatile and difficult to form expectations over. For this reason, DAOs contract in a mix of cash, stablecoins and native tokens. These different strategies influence the way DAOs attract, retain, and motivate talent.

The stronger the signal the DAO can send about its quality, and thereby the quality of its native token (if it has one), the better deal it can make in terms of compensation, meaning it can hire higher quality workers, further improving the quality of the project. There is a fly-wheel effect that is conditional upon effective signalling. DAOs share many challenges faced by startups: both are young, liquidity-constrained organisations that must attract, retain and motivate skilled knowledge workers to build external offerings and internal structures. Typically, they cannot pay competitive market rates for these workers, and face high levels of risk and long term uncertainty. Many people drawn to working with DAOs are motivated by non-pecuniary interests, such as a shared vision or sense of purpose, and are willing to accept high levels of risk and forgo immediate income. Both DAOs and traditional startups leverage these non-financial benefits to make up for their inability or unwillingness to offer higher salaries.

Consider the problem of hiring into a DAO. In a conventional organisation, a manager or committee will run through a process to attract, sort, screen, verify, and evaluate a candidate, make an offer, onboard, engage with HR and external stakeholders (e.g. unions) and then supervise work, often in a team. Almost none of these institutional processes are available in the same way in the process of working for a DAO (Ilyushina 2023). The challenges DAOs face in hiring are representative of broader problems they face in dealing with asymmetric information.

The decentralised and autonomous nature of DAOs are also sources of quality control and reputation problems. The traditional employment contexts that rely on established markers of quality — such as prestigious degrees, employment at reputable firms, face-to-face interviews and referee reports (Davila et al., 2003) — are difficult to transfer to DAOs that are based on decentralised and autonomous processes, and so require different methods for interpreting honest markets of ability. Furthermore, potential workers can be reluctant to participate in a market perceived as volatile or vulnerable to manipulation, and therefore discount the value they expect to gain from participation. Decentralisation and autonomy create conditions in which dishonest communication can become rife, both toward DAOs (malicious agents seeking to exploit through misrepresentation) or outward from DAOs (as possible sources of scams). DAOs must develop new institutions to solve these asymmetric information problems in their ecosystems.

## 4 New institutions for asymmetric information in DAOs

In this section we consider a range of emergent institutions, strategies and tools that DAOs, and the broader ecosystems in which they operate, have developed and evolved to solve these problems of asymmetric information about quality. DAOs have developed and deployed various specialist functions, such as DAO creation, discussion, communication and governance tools (see Table 1 in Appendix A). Each of these services and operations are distinct capabilities and services that have drawn in a range of specialist start-up firms to offer these services. With only a few exceptions, these services are offered by new companies or products (or related areas). Below we discuss several examples of mechanisms that mitigate asymmetric information in DAOs through costly signalling.

### 4.1 Institutions as products

Here we outline three products that have emerged as costly signalling mechanism in DAO communities. *Dework* is a decentralised employment marketplace where any DAO can list public bounties in exchange for payment or rewards and receive expressions of interest. If matched, upon completion, the DAO will automatically award the bounty and a non-fungible token (NFT) proof of completion. This NFT is a receipt for a digital wallet, including details of the bounty (task name, DAO, assignee, reviewer). This POAP is an immutable reputation token and inter-organisational signal that proves a DAO worked with an individual.

*Coordinape* helps distribute wages through a community voting process. Unlike bounty hunting, working group members pool efforts for a regular (monthly) income. Actions and payments performed by the DAO are visible to all through the protocol, allowing internal voting and redistribution (through kudos). Coordinape is intended to ensure fair distribution of wages by collectively determining who contributes value and deserves reward. The DAO can gain valuable insights about work prioritisation, community value, and key contributors in various areas.

*Sourcecred* connects to a DAO's existing social platforms, typically Discord and Discourse, and employs various metrics such as engagement, length of membership, and continued activity to determine a 'cred' score for each community member that can be converted into benefits that include privileged access to internal bounties and ability to monetise cred. Sourcecred score serves as a costly signal that facilitates entry or access to specific work opportunities (Rennie 2023).

### 4.2   Emergent phenomena and behaviour

Rather than the use of specific products (as is the case of Dework, Coordinape and Sourcecred) there are several other mechanisms by which the costly signalling problem in DAOs is mitigated. While a large and unspent DAO treasury, for instance, has a clear utilitarian purpose to fund the development of local public goods for the DAO ecosystem, it is also a costly signal about the quality of a project and a community (Allen et al. 2021). A strong and high quality DAO can leave vast capital reserves in a treasury or a foundation, including baring the exchange rate risk of holding their native token. A large treasury creates a costly exit moat in terms of opportunity cost, because any players seeking to fork the project can't take the DAO treasury with them. That stabilises expectations about community stickiness.

Web3 cultural literacy is also a costly signal of investment in web3 communities. The widespread use of what are often called 'dank memes' and emojis can act as a costly signal of investments in a community. Workers in DAO labour markets are often pseudonymous (with persistent online identities) and interact with current and potential future employers in online channels. This creates an incentive to display evidence of high activity and in-group status in these spaces. Performative activity (e.g. saying 'gm', short for 'good morning', or particular emoji posting, e.g. Laser-eyes, or high cache PFPs, such as pudgy penguins or BAYC) or fluency with niche memes are signals of group affinity (as well as 'proof of humanity') are all costly signals. Separating equilibrium theory predicts rapid cultural evolution in, i.e. the 'dankness' or freshness of the meme. Note the 'costlinesss' here is the risk of making easily observable cultural mistakes that can cause reputational harm of being a pretender or fake participant (i.e. 'cringe').

## 5      Costly signals in DAOs

DAOs face an asymmetric information problem of communicating information about their true state. Any DAO can claim an excellent codebase, great and trustworthy founding team, good intentions and a worthy mission. Sending costly signals are ways for true high-type DAOs to distinguish themselves from low-type DAOs that might be seeking to pool with high-type DAOs to create a mixed and noisy signal. This pooling equilibria is a problem for high-type DAOs, because counterparties will form expectations over the pooling equilibrium and make bids or offers that are beneficial for low-type DAOs but poor outcomes for high-type DAOs. High-type DAOs are thus incentivised to invest in separating equilibrium signals that are easily evaluated by counterparties (Bergh et al. 2014). Industrial era organisations have long adapted to this problem, engaging in efficient costly signalling and with centuries-old collective action solutions to facilitate signalling in the form of local public goods and institutions, such as regulation, licensing and credentials. But DAOs are a radically new organisational and institutional technology, and so the ecosystem of supporting institutions (Jacobides et al. 2018) that might facilitate such signalling is highly underdeveloped.

There are two challenges that DAOs must overcome: (1) figuring out good and effective signalling investments and strategies in this new environment, and to these new counterparties; and (2) ensuring the signalling mechanisms to send these signals get built and work effectively. The first challenge each DAO, or agent seeking to engage with a DAO, must solve individually. The second challenge is both an entrepreneurial opportunity (to build signalling mechanisms, and to capture some of the signalling rents) and a collective action problem at the level of the DAO ecosystem.

The examples in section 4 above illustrate a mix of signalling mechanisms that have developed as entrepreneurial opportunities as collective action solutions to build signalling infrastructure and institutions across the DAO ecosystem. This is in line with economic theory predictions, which emphasise that there are potentially substantial rents to be had by solving this broad information problem to lower the transaction costs associated with economic interaction with DAOs. Signalling behaviours require the pre-existence of signalling mechanisms and institutions. These must be built-up and developed for DAO ecologies to be able to effectively compete with the mature signalling ecologies of industrial era organisations and labour markets.

An important question is who bears the cost of investing in the creation of these institutions and infrastructure. A purely private ordering solution, where each agent both invests in the signal and in the signalling mechanism, is likely to be inefficient if there are fixed costs and uncompensated externalities associated with the mechanism, which is surely the case. But a pure public ordering solution (to a local public goods problem) is also unlikely to be optional due to the substantial information and coordination problems in overcoming free-rider effects, as well as the rent protection actions to raise the cost of such mechanisms from those who benefit from the industrial-era signalling mechanisms incumbent in politics (Juma 2016). While the costs of discovery of effective signals are likely to be internalised by those who discover and therefore first benefit from them, these strategies will likely be copied by others. However, the costs of building new signalling mechanisms to discover and exploit such signals is an externality distributed over a range of potential beneficiaries, including existing and prospective community members, current and new investors, regulators and legislators, and prospective workers or contributors. While each stakeholder has individual incentive to build and contribute to the mechanism (i.e. private provision is possible), each is also better off if someone else does – so a collective action problem lurks at the core of this ecology. We need to study how these common pool resource problems are solved, where solutions do emerge. A promising way of approaching this has been developed as contribution systems (Keally and Ricketts 2014, Potts and Rennie 2023) that make use of the same algorithmic tools (consensus mechanisms, token-gated participation, distributed voting and payments, etc) that have caused the problems in the first place.

We can also witness the emergence of countersignalling equilibria. Countersignalling are nonmonotonic separating equilibria in noisy signalling environments between low, medium, and high types (Feltovich et al. 2002). High-types are not worried about being confused with low-types, while medium types are. Medium types send costly separating equilibria signals, and high-types send pooling equilibria signals with low types (i.e. they countersignal). Conspicuous consumption is a costly signal of wealth that creates a separating equilibrium with the poor (Sundie et al. 2011), and investment in compliance with regulations and licensing regimes is a costly signal of quality for an organisation (like an education credential for a worker) to differentiate from lower quality organisations that cannot economically meet the threshold. But under some circumstances conspicuous law-breaking or refusal to comply with regulatory mandates (and therefore pooling with the low-types) also can be a costly signal of power, resources and determination to create countersignalling separating equilibrium when there are for instance, powerful network effects in play, as is the case in all platform businesses. Uber for instance used this strategy in defiance of local taxi licensing regimes, as a costly signal to drivers to join the platform. This predicts that protocol DAOs, such as DEXs, will likely use such countersignalling strategies when competing for winner-take-most global infrastructure opportunities. Alternatively, conspicuous and performative law-following, even at great cost (e.g. when platforms actively seeks financial licensing compliance in the US jurisdictions) as a costly signal to consumers worldwide.

## 6 Conclusion

The well-known problems of fraud and scams in DAOs can be reframed as a problem of ineffective costly signalling mechanisms. Without those costly signalling mechanisms, high-quality DAOs cannot easily differentiate themselves from low-quality DAOs. The development of new and better signalling mechanisms will help the DAO ecosystem to develop. Yet the success of any individual DAO will partly depend on the ability as a collective to develop better solutions to signalling problems. Solving this involves collective action to internalise the externality. High-type DAOs are powerfully incentivised to find private order solutions to the signalling problem, including making investments in signalling mechanisms, some of which can be local public goods (at the level of DAOs as a legal type, or at the level of particular industries in which they operate, such as DeFi). Signalling mechanism infrastructure can be protocols for regulation management and exchange, or licensing and credentialing regimes, etc, and may or may not require public support (e.g. regulation). We have shown here that the development of effective signals is itself an experimental discovery process that is being driven by entrepreneurial agents seeking to provide solutions, often in the form of aspects of DAO tooling or infrastructure. This process is happening on two sides: (1) as DAOs seek to send signals about their true type to counterparties (e.g. members, investors and regulators); and (2) as workers seek to signal to DAOs their hidden qualities. We anticipate the emergence of further costly signalling mechanisms as the industry progresses, including mechanisms that leverage the unique data and environment of Web3.

# Appendix

**Table 1.** DAO services and the operations that provide them

| Tool | Examples |
| --- | --- |
| DAO creation tools | Aragon, DAOstack, Colony, Syndicate, OpenLaw, DAOhaus |
| Discussion tools | Discord, Discourse, Telegram |
| Operations, tasks, recruiting | Dework, Notion, Trello, Metropolis, SourceCred, Rabbithole |
| Token gating | CollabLand, Matrica, Guild |
| Governance & voting | Snapshot, Tally, Boardroom |
| Treasury management | Gnosis Safe, Llama, Utopia |
| Finance & payments | Coordinape, Superfluid |
| Front-end & analytics | DeepDAO, DAOHQ, Messari |
| Ecosystem support | MolochDAO, Gitcoin |

**Table 2.** DAO functions that deal with asymmetric information problems

| **Foundational Infrastructure** | |
|---|---|
| Decentralised storage | storage system that distributes data across multiple nodes within a network, enhancing security and reliability. e.g. IPFS. signals robustness. |
| Decentralised identity | Proof of humanity (https://proofofhumanity.id/). Soul bound tokens: digital token that remains permanently associated with an individual's wallet that cannot be transferred and used to authenticate a person's credentials in applications (Weyl et al. 2022) |
| Proof of Attendance Protocol (POAP) | digital certificate that confirms and validates an individual's presence at a specific event, verifying claims and creating separating equilibrium with cheap talk (e.g. POAP.xyz) |
| Shared security | Interchain security, which is both an engineering feature but also additionally signals integration and cooperation with other projects. (Allen et al. 2023a) |
| Airdrops | Rewards to user contributions, signalling engagement by users, and also signalling commitment to users by protocols (Allen et al. 2023b). |
| Local public goods | MolochDAO (Soleimani, et al., 2019), RadicalxChange (matching funding), Gitcoin, (Buterin et al. 2019) |
| **Creating and managing credentials, engagement and contributions** | |
| Decentralised Work History Applications | enable individuals to establish and manage their work history in a decentralised manager, with privacy and control over personal data. e.g. DeWork, Dock |
| Contribution systems | platforms facilitate the decentralised distribution of compensation to individuals based on their contributions to specific projects or communities. e.g. Coordinape, Protocol Guild, SourceCred (Rennie 2023) |
| Reputation | Once an individual can prove their work history with a DAO, the quality of the DAO itself is a signal that candidates can leverage to showcase their skills |
| **Organisational and governance tools** | |
| Guilds | Individuals who collaborate into groups to share skills and expertise. Membership may require specific experience or proof of contributions. E.g. Metropolis ('pods', specialist groups within DAOs, with NFT-gated voting and access rights, Tracheopteryx, 2021), and Protocol Guild, a collective funding mechanism for core ethereum developers (https://protocol-guild.readthedocs.io) |
| Stewards | Developed by the Token Engineering Commons, stewards are tasked with governing platforms or communities. They are selected based on experience, skills, and contributions. Their appointment creates a signal of competence within the organisation, enhancing trust and collaboration. (Nabben 2021) |